# A possible cosmological effect on the quantum-to-classical transition

C. L. Herzenberg

*Herzenberg Associates, 1700 E. 56$^{th}$ Street #2707, Chicago IL 60637-5092*

Although cosmic expansion at very small distances is usually dismissed as entirely inconsequential, these extraordinarily small effects may in fact have a real and significant influence on our world. A calculation suggests that the minute recessional velocities associated with regions encompassed by extended bodies may have a role in creating the distinction between quantum and classical behavior. Using the criterion that the uncertainty in position should be smaller than the size of an object together with estimates based on the range of Hubble velocities extending through the object lead to a threshold size that could provide a fundamental limit distinguishing the realm of objects governed by classical laws from those governed by quantum mechanics.

The question of how and under what circumstances quantum mechanical behavior can become effectively classical is of significance for the foundations of quantum mechanics and also for practical applications. Issues relating to the approach to the classical limit and to quantum measurement have been examined for many years without an entirely satisfactory resolution. What are the conditions under which classical physics will emerge as an approximation to quantum physics? Independently of whether, for example, this may be an issue of sufficiently large quantum numbers, or whether classical physics may exist independently of quantum mechanics and might not be derivable from it, we may presumably anticipate the existence of some basic criteria distinguishing the conditions of quantum behavior of microscopic objects from the conditions associated with the classical behavior of macroscopic objects.

One approach to the quantum measurement paradox has been the concept that quantum mechanics is in the final analysis not a complete description of the physical world, but must be supplemented by other laws or other aspects of physics. Since the predictions of quantum mechanics are extremely well verified at the atomic level, any corrections due to such extraneous laws or effects must be very small in the atomic regime while still sufficiently effective at the macroscopic level. We propose here such a candidate effect, based on the Hubble expansion of space.

Environment-induced quantum decoherence has been considered the key process by which quantum systems in complex environments exhibit classical behavior, and can be responsible for the transition from a quantum to a quasi-classical state [1] [2] [3] [4] [5] [6]. Quantum decoherence occurs when a system interacts with its environment in such a way that the system loses phase coherence between different portions of its quantum mechanical state, and it is usually discussed in terms of coupling to an environment as a result of which wave functions for macroscopically distinct states rapidly become orthogonal, as a consequence of the dense energy spectrum of the environment and the extremely high dimensionality of the relevant Hilbert space. The key idea promoted by the approach of quantum decoherence is the insight that realistic quantum systems are never isolated, but are immersed in the surrounding environment and interact continuously with it [3].

We will examine the intrinsic immersion of a quantum object in a space that exhibits continuous expansion, and we will examine the effects of the continuous interaction of the object with this more basic environment on the quantum behavior of such an object. The presence of the intrinsic cosmic expansion would cause objects or parts of objects at separate locations to be in relative motion with respect to each other, and thus could potentially affect quantum wave functions over distances characterizing macroscopic objects.

While cosmological expansion, or Hubble expansion, is customarily discussed in conjunction with cosmic distances, and particularly for gravitational systems not bound to each other, the possibility of effects associated with an underlying expansion at smaller distances should not be overlooked. Thus Hubble expansion is generally regarded as originating from the expansion of the universe, and to be a characteristic of space itself. The existence of Hubble expansion on cosmological scales has been very well established for many years, and there has been some revival of interest in whether it affects phenomena at distances smaller than cosmological distances [7] [8] [9] [10] [11]. Hubble expansion does not appear present in gravitationally bound systems such as galaxies, a suppression in conformity with the virial theorem. The question has been discussed as to whether there exists any cut-off distance such that systems below that scale do not participate in the Hubble expansion, and, if so, what could possibly determine such a maximum scale [7] [9]. However, the cosmological metric alone does not dictate a scale for expansion, and, furthermore, it is difficult to justify any particular scale for the onset of expansion or the 'shielding' of systems smaller than that scale from cosmic expansion [7]. Evidence for Hubble expansion has been sought at distances less than cosmic distances, and the anomalous acceleration of the Pioneer spacecraft has been interpreted as a possible manifestation of Hubble expansion within the solar system [11]. Furthermore, some evidence for residual Hubble expansion effects has been reported from lunar laser ranging measurements [12] [13]. What about even smaller scales? It is usually assumed that local systems such as the molecular or atomic structure are unaffected by the cosmic expansion [7]. While the effect is actually registered over the small distances of the wavelengths of light, this is generally regarded as an imprint of the expansion at larger scales [9]. But in principle Hubble expansion could be present at the smallest practical scale and observable in principle [9]. Other authors have argued that there is only one space-time and that all physical systems, large and small, will feel the effects of this cosmic expansion in one way or another [7]. In order to examine the potential implications, we will assume that Hubble expansion is present at the smallest scales and observable in principle, and that every point in space is a locus of expansion.

We will examine the possibility that the extremely small relative velocities intrinsically present at different locations within the space occupied by a macroscopic object could contribute to bringing about a quantum-to-classical transition.

We consider an extended object that consists of an assemblage of components atoms or parts located in proximity to each other within a defined space. As a consequence of universal expansion, different regions of this object may be regarded as potentially moving apart from each other at extremely slow rates, with velocities of recession linearly dependent on their distance apart, since the cosmic recessional velocity between two locations is proportional to the distance separating the locations. Thus, because of Hubble expansion, any extended object would exhibit a range or spread in velocities with an approximate value:

$$\Delta v \approx H_o L \qquad (1)$$

Here, $H_o$ is the Hubble constant, L is a length characterizing the size of the object, and $\Delta v$ is the spread in recessional velocities within the object. We will treat this intrinsic spread in velocities that would necessarily characterize an extended object as providing a measure of an uncertainty in velocity associated with the object as a whole.

We can calculate the associated uncertainty in momentum by forming the product of the mass of the object with the uncertainty in velocity associated with the object:

$$\Delta p = m \, \Delta v \approx m H_o L \qquad (2)$$

Employing the Heisenberg uncertainty principle which states that $\Delta p \, \Delta x \geq h/4\pi$, where h is Planck's constant, we can evaluate an approximate limit on an associated uncertainty in position as:

$$\Delta x \geq h/(4\pi m H_o L) \qquad (3)$$

Thus, we see that for larger and more massive objects, the uncertainty in position associated with this effect decreases, whereas for smaller objects the uncertainty in position associated with this spread in velocities will increase.

For simplicity we will treat the object as roughly cubical; then this relationship can be expressed in terms of the density approximately as:

$$\Delta x \geq h/(4\pi \rho H_o L^4) \qquad (4)$$

To examine the critical limiting case or threshold value of the minimum uncertainty in position being comparable to the linear dimension of the object, we can set the value of the uncertainty $\Delta x$ equal to the length of the object and obtain as an estimate for the critical value of the length $L_{cr}$:

$$L_{cr} \approx [h/(4\pi \rho H_o)]^{1/5} \qquad (5)$$

If we insert numerical values for the parameters (using a choice of density 1 gram per cc to roughly characterize macroscopic objects, and a Hubble constant of $2.3 \times 10^{-18}$ per second), we find that $L_{cr}$ can be evaluated as of the order of magnitude of 0.1 millimeter.

Thus, this criterion would suggest that objects of sizes greater than about 0.1 millimeter or masses greater than about a microgram would be expected to behave in a classical manner, while objects of appreciably smaller sizes and smaller masses could exhibit quantum behavior as entire objects.

It may be noted that using a slightly modified alternative approach of modeling such an object as a wave packet would appear to lead to a similar result. Because of the Hubble recession, different portions of the mass of the object will be characterized by different recessional velocities. A wave packet would be formed from the superposition of wave functions describing different portions of the object, and these contributing wave functions would have wave numbers and hence wavelengths determined by the momenta associated with different parts of the object. The differences in values of the momenta of the participating wave functions would be roughly of the order of magnitude of the product of the objects mass with the maximum velocity difference within its extent, and hence approximately given by Eqn. (2). When the spread of wavelengths for the wave functions comprising the wave packet is compared to the object's characteristic length, we will again obtain results similar to those obtained above in Eqn. (5).

These results would appear to have a possible bearing on why microscopic objects are usually found in energy eigenstates, whereas macroscopic objects exhibit a spread of energies and occur in time-dependent states. But what are we to make of these threshold values? Are they useful? Perhaps not, as they clearly do not provide stringent limits. There is evidence that nearly all molecules except the very small ones seem to have reasonably well-defined spatial structures, and furthermore, larger biologically important molecules like alanine appear only in specific chiral configurations, although eigenstates of a parity-conserving Hamiltonian are symmetric or antisymmetric under parity transformation and then except in cases of degeneracy could not be chiral [1]. Thus observations seem to indicate that effective classicity can extend down to much smaller sizes than the limits that we have obtained, to the level of medium-sized molecules, presumably as a consequence of quantum decoherence from interactions with the environment. However, the present approach does appear to provide what may perhaps be a fundamental limit for a quantum-to-classical boundary. Furthermore, these values can provide limiting estimates above which quantum effects would be expected only within bound systems, rather than characterizing the behavior of an independent object as a whole, and below which quantum behavior may be present in appropriate circumstances. Thus, it appears that we may be able to consider cosmic expansion as setting a limit on the size of independent objects above which classical behavior may be expected to set in.

--------